# The Host Galaxies of Pulsar Timing Array Sources: Converting Supermassive Black Hole Binary Parameters into Electromagnetic Observables

Niccolò Veronesi[1], Maria Charisi[1,2], Stephen R. Taylor[3], Jessie Runnoe[3,4], and Daniel J. D'Orazio[5,6]
[1] Department of Physics and Astronomy, Washington State University, 1245 Webster Hall, Pullman, WA 99164, USA; veronesi.nicco@gmail.com
[2] Institute of Astrophysics, FORTH, GR-71110, Heraklion, Greece
[3] Department of Physics and Astronomy, Vanderbilt University, 2301 Vanderbilt Place, Nashville, TN 37235, USA
[4] Department of Life and Physical Sciences, Fisk University, 1000 17th Avenue N, Nashville, TN 37208, USA
[5] Space Telescope Science Institute, 3700 San Martin Dr., Baltimore, MD 21218, USA
[6] Department of Physics and Astronomy, Johns Hopkins University, 3400 North Charles Street, Baltimore, MD 21218, USA

## Abstract

Pulsar timing arrays (PTAs) are approaching the sensitivity required to resolve gravitational waves (GWs) from individual supermassive black hole (SMBH) binaries. However, the large uncertainty in source localization will make the identification of its host environment challenging. We show how to convert the posterior probability function of binary parameters inferred by GW analyses into distributions of apparent magnitudes of the host galaxy. We do so for a scenario in which the host environment is a regular early-type galaxy, and one in which it is an active galactic nucleus. We estimate the reach of PTAs in the near and intermediate future, and estimate whether the binary hosts will be detectable in all-sky electromagnetic (EM) surveys. A PTA with a baseline of 20 yr and 116 pulsars, resembling the upcoming data release of the International Pulsar Timing Array, can detect binaries out to a luminosity distance of 2 Gpc (corresponding to a redshift of $z \sim 0.36$), while a PTA with a baseline of 30 yr and 200 pulsars can reach out to distances slightly greater than 3 Gpc ($z \sim 0.53$). We find that the host galaxies of all binaries detectable with a baseline of 20 yr are expected to be present in the Wide-field Infrared Survey Explorer and SuperCOSMOS surveys, if they lie outside the plane of the Milky Way. The Two Micron All Sky Survey becomes incomplete for hosts of binaries more massive than $10^{9.8} M_\odot$ at a luminosity distance greater than 1 Gpc. The EM surveys become slightly more incomplete when PTAs with longer baselines and therefore improved sensitivities are considered.

## 1. Introduction

Black holes with masses greater than $10^6 M_\odot$ occupy the centers of most (if not all) massive galaxies (D. Richstone et al. 1998). The masses of these supermassive black holes (SMBHs) are correlated with several large-scale properties of the host environments, such as the velocity dispersion and the mass of their bulge (L. Ferrarese & D. Merritt 2000; K. Gebhardt et al. 2000; N. J. McConnell & C.-P. Ma 2013). This empirical correlation suggests that galaxies and their central SMBHs coevolve through cosmic history (J. Kormendy & L. C. Ho 2013).

In hierarchical structure formation, SMBH binaries are expected to form as a consequence of galaxy mergers (M. C. Begelman et al. 1980). If binaries are hardened enough to reach subparsec separations, they emit low-frequency gravitational waves (GWs). Binaries with masses of the order $10^3$–$10^8 M_\odot$ emit millihertz GWs and will be detectable with the Laser Interferometer Space Antenna (P. Amaro-Seoane et al. 2023), while binaries with total masses greater than $10^8 M_\odot$ emit nanohertz GWs and are detectable by pulsar timing arrays (PTAs).

PTAs use observations of multiple pulsars in the radio band, monitoring the time of arrival of their pulses in order to detect the deformations in the spacetime structure that low-frequency GWs consist of. PTAs have been operating for approximately 20 yr, and they rely on precise measurements of dozens of stable millisecond pulsars across the Galaxy. GWs are detected by measuring inter-pulsar correlations in the difference between predicted and observed times of arrival of pulsar signals (i.e., residuals; M. V. Sazhin 1978; S. Burke-Spolaor et al. 2019; S. R. Taylor 2021).

Recently, all major PTA collaborations, i.e., the North American Nanohertz Observatory for Gravitational Waves (NANOGrav; M. A. McLaughlin 2013; S. Ransom et al. 2019), the European Pulsar Timing Array (EPTA; M. Kramer & D. J. Champion 2013) in combination with data from the Indian Pulsar Timing Array (InPTA; B. C. Joshi et al. 2018), the Parkes Pulsar Timing Array (PPTA; R. N. Manchester 2008; G. Hobbs 2013), the Chinese Pulsar Timing Array (CPTA; K. J. Lee 2016), and the Meerkat Pulsar Timing Array (MPTA; M. Bailes et al. 2020) have found evidence for the presence of a GW background in the nanohertz band (G. Agazie et al. 2023b; EPTA Collaboration et al. 2023; D. J. Reardon et al. 2023; H. Xu et al. 2023; M. T. Miles et al. 2025). The levels of significance vary among different PTAs, but the inferred parameters of the GW background (amplitude and spectral shape) are broadly consistent (G. Agazie et al. 2024). The measured signal is likely produced by the superposition of GWs coming from a population of unresolved SMBH binaries (G. Agazie et al. 2023c; EPTA Collaboration et al. 2024).

Thanks to the continuously improving sensitivity of PTAs, the first resolved SMBH binary is expected to be detected in the next 5 to 10 yr (C. M. F. Mingarelli et al. 2017; L. Z. Kelley et al. 2018; B. Bécsy et al. 2022; E. C. Gardiner et al. 2024). However, the localization uncertainty of the source is expected to be of the order $10^2$–$10^3$ square degrees, and the luminosity distance from the Earth will also be fairly unconstrained (J. M. Goldstein et al. 2019; P. Petrov et al. 2024). For this reason, a confident identification of the host galaxy through a selection solely based on sky position and luminosity distance is highly unlikely.

Several electromagnetic (EM) signatures, especially in the scenario of the host environment being an active galactic nucleus (AGN), have been proposed as potential indicators of the presence of a SMBH binary (see T. Bogdanović et al. 2022; D. J. D'Orazio et al. 2023 for reviews on the observable properties of SMBH binaries). After a preliminary galaxy selection based, for example, on their position has been done, searches in archival data or dedicated follow-up observations will be needed to identify such promising EM signatures in the host environment. These features include, but are not limited to, (1) resolving the system through direct imaging (D. J. D'Orazio et al. 2018), (2) Doppler shifts of the broad emission lines (T. Bogdanović et al. 2009; M. Eracleous et al. 2012; K. Nguyen & T. Bogdanović 2016; H. Guo et al. 2019), (3) periodic variability (Z. Haiman et al. 2009; D. Lai & D. J. Muñoz 2023), (4) radio jets with atypical (wiggly or helical) morphology (E. Kun et al. 2015; S. J. Qian et al. 2018), (5) X-ray spectra with atypical features like notches, soft X-ray excesses or oscillations in the Fe K-$\alpha$ line (B. McKernan et al. 2013; C. Roedig et al. 2014; M. L. Saade et al. 2020), (6) variations of the position of the photocenter in quasar precise astrometry (L. ć. Popović et al. 2012), and (7) morphological (J. Bardati et al. 2024a), kinematic, or chemical properties of the host, like slow rotation and strong misalignments of the stellar component (J. Bardati et al. 2024b; P. Horlaville et al. 2025), and high levels of metallicity (K. Cella et al. 2025). While most of these features are observable only in AGN, a recent theoretical work suggests that most of the hosts of PTA sources are unlikely to be associated with this type of host (R. J. Truant et al. 2025).

Since no confirmed subparsec SMBH binaries have so far been identified and there are uncertainties regarding the nature of their EM signatures, the first detection of a resolved PTA source will likely trigger follow-up observations or archival searches for the most promising potential counterparts. Following a completely agnostic approach and collecting enough evidence in favor for or against the presence of a SMBH binary for each of the thousands of galaxies within the typical localization area of a GW signal is highly demanding in terms of observational resources and computational time. Therefore, it is of pivotal importance to develop an efficient pipeline to first select and rank the most promising host galaxies based, for example, on their sky position, total mass, and luminosity distance (J. M. Goldstein et al. 2019; P. Petrov et al. 2024; R. J. Truant et al. 2025), and then trigger targeted searches only for a limited number of galaxies to look for, ideally, multiple EM signatures of a SMBH binary.

Aside from the position of the galaxies relative to the GW localization, their brightness in different bands of the EM spectrum can also be used to inform a ranking system of potential SMBH binary hosts. This is because the mass of the central SMBH (and thus likely the total mass of the hypothetical SMBH binary) correlates with the global properties of the host galaxy. Therefore, one can convert its posterior distribution, along with that of the luminosity distance, both obtained from the GW analysis, into a probability distribution of the expected apparent magnitude or flux of the host galaxy. This can also inform whether the galaxy catalogs (such as the one compiled in Z. Arzoumanian et al. 2021 and used in P. Petrov et al. 2024) used for the initial sky position, mass, and distance cuts are complete, and to what extent, within the region of the Universe that can be probed with PTAs. In this work, we present a method to convert the SMBH binary properties into the expected brightness of the host environment using different galaxy–SMBH mass scaling relations, while keeping track of the associated uncertainties.

We determine the range of binary parameters that PTAs may detect in the near and intermediate future, by simulating mock PTA data sets starting from the upcoming data set of the International Pulsar Timing Array (IPTA; G. Hobbs et al. 2010; J. P. W. Verbiest et al. 2016; B. B. P. Perera et al. 2019), which will combine public data from NANOGrav, EPTA+InPTA, PPTA, and MPTA with a baseline of $\sim$20 yr. We also extend our analysis to baselines of 25 and 30 yr, progressively adding more pulsars. We use these simulations to calculate the maximum luminosity distance (or minimum mass) of SMBH binaries that can produce detectable GWs with the upcoming PTA sensitivity, and we estimate whether or not recent all-sky EM surveys can include the potential hosts. Each analysis presented in this work is done assuming two different scenarios: (a) the host galaxy is a regular early-type galaxy (ETG), and (b) the host galaxy has an AGN.

The paper is structured as follows. In Section 2, we describe how we generate our mock IPTA data sets and the binary detection method. The method used to convert SMBH binary properties into probability distributions of the brightness of the host environment is detailed in Section 3. In Section 4, we present our results, while in Section 5, we discuss the implications of our findings, some caveats to our analysis, and future steps. We present our summary and conclusions in Section 6.

## 2. Data Sets

In this section, we describe the mock IPTA data sets we use in our analysis and the method adopted to create them.

The analysis presented in this work uses three different configurations of PTAs with increasing baselines and number of monitored pulsars. They are set up to represent realistic scenarios of future IPTA data sets, with three different baselines of 20, 25, and 30 yr. We will further refer to these configurations as IPTA_20, IPTA_25, and IPTA_30, respectively.

### 2.1. Configuration of the Pulsar Arrays

The IPTA_20 data set has the same configuration as the one detailed in Section 2.3.1 of P. Petrov et al. (2024), and resembles what we expect for the upcoming third data release of IPTA. It is composed of 116 pulsars in total, 68 of which are from the NANOGrav 15 yr data set (G. Agazie et al. 2023a), 31 from the first data release of MPTA (M. T. Miles et al. 2023), 14 from the third data release of PPTA (A. Zic et al. 2023), and three from EPTA+InPTA DR2new+ (EPTA Collaboration et al. 2023). We note that some of the pulsars are monitored by multiple PTAs (e.g., see Figure 9 in G. Agazie et al. 2024), and in that case we choose the NANOGrav observations as the basis for our simulations.

For the 45 pulsars included in the NANOGrav 12.5 yr data set, we reach a total baseline of 20 yr adding observations with the same procedure followed in P. Petrov et al. (2024) and N. S. Pol et al. (2021). Specifically, the cadence of simulated observations is drawn from the distribution of observed cadences in the last year of data collection for each pulsar, excluding observations more frequent than one per day. Similarly, the uncertainties on the times of arrival are sampled from the distribution of such uncertainties over the last year of observations. For the remaining 71 pulsars, which were not part of the NANOGrav 12.5 yr data set (M. F. Alam et al. 2021), the timing model parameters are randomly sampled from the following four pulsars: J0931−1902, J1453+1902, J1832-0836, and J1911+1347 (see P. Petrov et al. 2024 for details on the choice of those pulsars). For those, we reach a total baseline of 20 yr by adding one observation every 2 weeks. In this case, the uncertainties on the times of arrival are taken from the white-noise values for each pulsar in the respective data set paper.

The simulated IPTA_25 and IPTA_30 data sets are composed of a total of 158 and 200 pulsars, respectively. We add seven new pulsars for each new simulated year of data collection. Following standard practices for the addition of new pulsars in a data set, we assume that the newly added pulsars have been monitored for at least 3 yr. The sky positions of the new pulsars are randomly drawn from the 2D probability density function of the equatorial coordinates of the 116 pulsars in IPTA_20, estimated with kernel density estimation (KDE). The timing model parameters for these new pulsars are again sampled from the four pulsars mentioned above. The sky positions of all the pulsars that comprise the different PTA data sets are shown in Figure 1. With blue stars we show the positions of the 116 pulsars in IPTA_20, while the yellow circles and white squares show the new pulsars added in IPTA_25 and IPTA_30, respectively.

### 2.2. Determination of the Best and Worst Sky Positions

Since the distribution of pulsars on the sky is highly anisotropic, the sensitivity of PTAs to individually resolved sources is similarly dependent on the sky position of the SMBH binary. In order to estimate the range of binary parameters that can be detected in the three data sets we consider, we first need to determine the most sensitive and least sensitive sky location of each array composition. For this, we inject a binary signal with the same parameters in different sky locations and estimate the resulting signal-to-noise ratio (S/N), following the procedure in P. Petrov et al. (2024). The individual binaries are injected on top of an isotropic GW background with an amplitude of $A_{\mathrm{GWbackground}} = 6.4 \times 10^{-15}$ at a reference frequency of 1 $\mathrm{yr}^{-1}$ and a power-law exponent $\gamma = 3.2$. These values are adopted from G. Agazie et al. (2023b) and correspond to the best-fit values for the general model power law that allows for a variable exponent, not fixing it to the fiducial value of $\gamma = 13/3$ predicted for a population of SMBH binaries with circular orbits whose evolution is determined by GW emission only. However, injecting a GW background with a slightly different power-law index is not expected to have any significant impact on our results.

We divide the sky in a HEALPix projection with a total of 192 pixels ($N_{\mathrm{Side}} = 4$) of equal area. We inject at the center of one of these pixels a reference SMBH binary with a total mass of $M_{\mathrm{tot}} = 10^9 M_{\odot}$, a mass ratio of $q = 1$, and luminosity distance from Earth of 100 Mpc. We set the binary orientation to face-on (i.e., its orbital inclination, defined as the angle between the angular momentum vector and the line of sight, is $\iota = 0$), and the GW frequency to $f = 10^{-8}$ Hz. We repeat this for all 192 pixels and for the three different PTA configurations we explore.

For each injected SMBH binary, we calculate the S/N as follows:

$$\mathrm{S/N} = \sqrt{\mathrm{s}^T \cdot C^{-1} \cdot \mathrm{s}}\,, \tag{1}$$

where $s$ is the vector containing the GW signal coming from the SMBH binary, $s^T$ its transpose, and $C$ is the noise covariance matrix that takes into account the intrinsic white and red noise of each pulsar and the GW background, which is modeled as a common uncorrelated red noise.

In Figure 1, we show the S/N as a function of sky position for the IPTA_30 data set. The upward and downward maroon triangles represent the best and worst sky positions for detecting individual SMBH binaries in IPTA_30, respectively, determined from the pixels that resulted in the highest and lowest S/N for the same injected binary. The best and worst sky locations have also been calculated for IPTA_20 and IPTA_25, shown with pink and orange triangles, respectively. The locations of the best and worst sky positions do not change significantly across different PTA configurations, with the best one always being in the sky region with the highest density of accurately timed pulsars (i.e., close to the Galactic plane), and the worst one approximately at the opposite side of the sky. Moreover, we calculate the pixel-by-pixel increase in the measured S/N as the baseline increases by 5 yr and 42 new pulsars are added in the array. We find an average increase of approximately 37% between IPTA_20 and IPTA_25 and of approximately 30% between IPTA_25 and IPTA_30.

### 2.3. Estimation of the PTA Reach

Once the best and worst sky locations are determined, we evaluate the reach of each PTA configuration to compare with the reach of different all-sky EM surveys. We calculate the binary parameter space that is accessible to each PTA by injecting binaries into the best and worst sky locations, and varying the total mass and the luminosity distance of the binary. For the injections, we assume circular orbits and the most optimistic scenario for the binary inclination, i.e., face-on orientation. As far as the mass ratio is concerned, we bracket the possibilities by examining equal-mass binaries with $q = 1$ and unequal-mass binaries with $q = 0.1$. We adopt a maximum value for the total mass of $10^{10} M_{\odot}$, motivated by SMBH binary population analyses which suggest that higher values are expected in at most 5% of the detectable systems (E. C. Gardiner et al. 2025).

We consider a SMBH binary to be resolved if its detection is characterized by $\mathrm{S/N} \geqslant 8$. This fiducial value is chosen because above this S/N threshold the localization area can in general be well constrained, like the other parameters of the binary (P. Petrov et al. 2024), even if the detection itself could be claimed with even lower values. This is important for this work, since we are interested in the identification of the binaries' host galaxies.

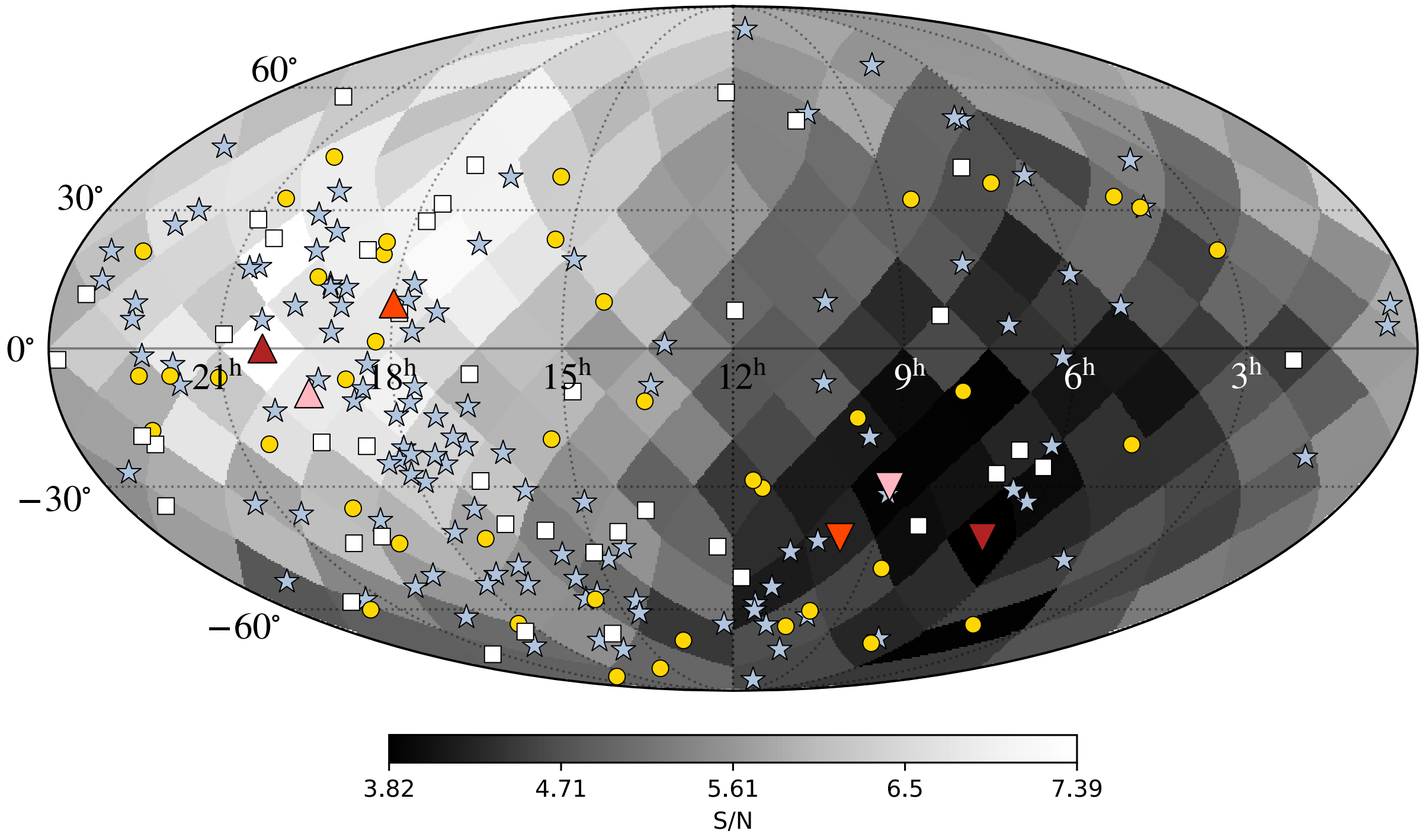

**Figure 1.** Mollweide projection of the signal-to-noise ratio (S/N) measured for a mock SMBH binary as a function of its sky position. The sky map shows the S/N measured for the IPTA_30 configuration. Stars represent pulsars in the IPTA_20 data set, circles mark the 42 pulsars added to create IPTA_25, and squares the additional 42 introduced in IPTA_30. The upward pink, orange, and maroon triangles mark the position of the pixel with the highest S/N in IPTA_20, IPTA_25, and IPTA_30, respectively. Analogously, the three downward triangles show the positions of the pixel with the lowest S/N for these data sets, following the same color-coding.

In Figure 2, we show the reach (i.e., the maximum distance at which a binary can be detected) of the three PTA configurations, evaluated in correspondence of three different values of binary total mass, $10^{9}M_{\odot}$, $10^{9.5}M_{\odot}$, and $10^{10}M_{\odot}$, and then interpolated. Different line colors correspond to different IPTA configurations. We consider three different GW frequencies, $10^{-8}$ Hz, $10^{-8.5}$ Hz, and $10^{-9}$ Hz, represented in the figure by solid, dashed, and dotted lines, respectively. Out of these three scenarios, the most optimistic one for binary detection corresponds to the signal with a frequency of $10^{-8}$ Hz. Note that we also explored frequencies of $10^{-7.5}$ Hz, which did not result in a significant increase in PTA reach with respect to the $10^{-8}$ Hz case. For clarity, in the figure we do not show this test. This choice is also in line with theoretical expectations which predict that only a small fraction of the expected SMBH binary detections is characterized by frequencies higher than 10 nHz (B. Bécsy et al. 2022; E. C. Gardiner et al. 2025). Given the significantly better sensitivity of our simulated PTAs in detecting GW signals at $10^{-8}$ Hz compared to the other values examined in this test, we choose to use this frequency in the rest of the analyses presented in this work.

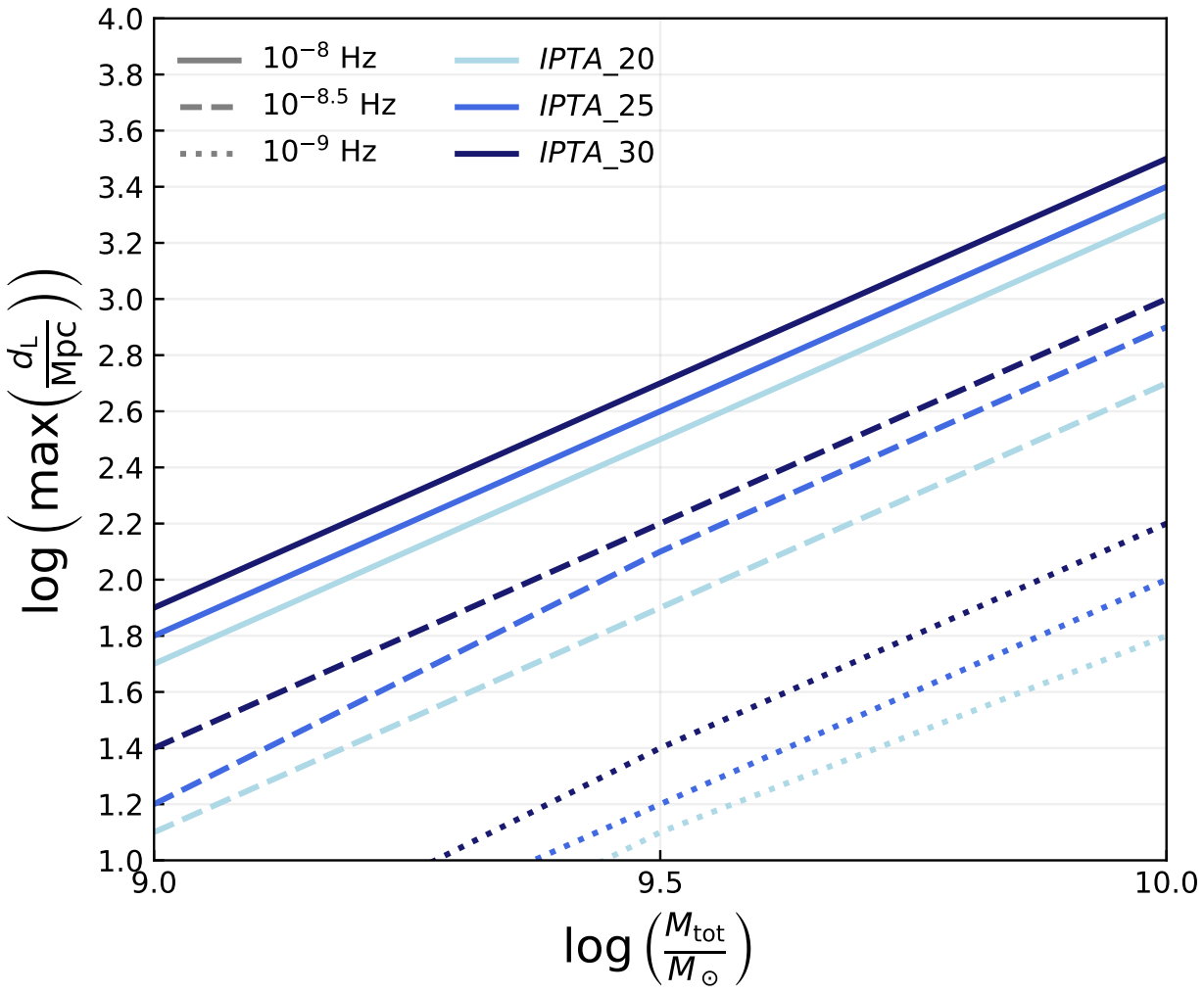

**Figure 2.** Maximum luminosity distance at which individual SMBH binaries can be detected by future IPTA data sets as a function of the binary total mass. Different shades of the lines indicate different baselines of the PTA data set, and different line styles indicate different frequencies.

## 3. Method

In this section, we present a method to estimate the EM properties, in particular the apparent magnitudes of the host galaxy of a SMBH binary, starting from the parameters of the latter as inferred from GW analysis. We take an agnostic approach regarding the type of galaxy with which the host environment is most likely to be associated. Hence, we calculate observed brightnesses for two cases: (1) when the binary resides in the center of a regular ETG, and (2) when the binary resides in an AGN. Recently, R. J. Truant et al. (2025) presented a detailed analysis on the theoretical expectations regarding the nature of the hosts of SMBH binaries and their apparent magnitudes. Our work here is complementary to that, examining the EM properties of the host galaxies from an observational perspective.

### 3.1. Conversion from SMBH Binary Parameters to ETG Apparent Magnitudes

In ETGs the mass of the central SMBH is correlated with the bulge mass, and thus with the stellar mass of the galaxy, which in turn is correlated with its luminosity in the near-infrared (near-IR) $K_s$ band. Therefore, the SMBH binary total mass and the luminosity distance can be used to estimate the apparent magnitudes of the host galaxy. This method was used to estimate SMBH masses from $K_s$-band luminosities in the galaxy catalog of Z. Arzoumanian et al. (2021). Here we use the same approach but in the opposite direction, i.e., we start from the binary total mass and estimate the $K_s$-band magnitudes.

Specifically, starting from the binary total mass, $M_{\rm tot}$, we use the $M_{\rm BH}$–$M_{\rm bulge}$ scaling relation to estimate the mass of the stellar bulge of the host galaxy, $M_{\rm bulge}$. Analogously to what was done in Z. Arzoumanian et al. (2021), even though the scaling relation correlates the mass of the central (single) SMBH of a galaxy with the bulge mass, we can instead use the estimates of the total mass of the binary system. This is justified by the fact that these values have been calculated from dynamical tracers, which are not expected to be able to distinguish a compact binary from a single object. We therefore estimate $M_{\rm bulge}$ by inverting the following equation:

$$\log\left(\frac{M_{\rm tot}}{M_\odot}\right) = a_1 + b_1 \cdot \log\left(\frac{M_{\rm bulge}}{10^{11} M_\odot}\right), \tag{2}$$

where $a_1 = 8.46 \pm 0.08$ and $b_1 = 1.05 \pm 0.11$. We adopt the values from N. J. McConnell & C.-P. Ma (2013), in which the correlation has been measured to have an intrinsic scatter of $\sigma_{\rm intr,1} = 0.34$ (see Section 5 for a discussion on the possibility of using different best-fit values for the parameters of the $M_{\rm BH}$–$M_{\rm bulge}$ relation).

For ETGs, we assume that the mass of the bulge corresponds to the total stellar mass of the galaxy in the entire redshift range to which PTAs are sensitive (B. P. Holden et al. 2009). We then use such a stellar mass of the galaxy, $M_*$, to estimate its absolute magnitude in the $K_s$ band, $M_{\rm K_s}$, by inverting the following relation taken from M. Cappellari (2013):

$$\log\left(\frac{\rm M_*}{\rm M_\odot}\right) = a_2 + b_2 \cdot (M_{\rm Ks} + 23), \tag{3}$$

where $a_2 = 10.5829 \pm 0.0086$ and $b_2 = -0.4449 \pm 0.0091$. We also take into account that this correlation for the estimation of the absolute magnitude $M_{\rm K_s}$ has an intrinsic scatter of $\sigma_{\rm intr,2} = 0.14$.

The apparent $K_s$-band magnitude, $m_{\rm K_s}$, is then calculated by inverting the following standard relation, which also depends on the luminosity distance from Earth, $d_{\rm L}$, of the binary and thus of the galaxy that hosts it:

$$M_{\rm Ks} = m_{\rm Ks} - 5\log\left(\frac{d_{\rm L}}{\rm Mpc}\right) - 25 - 0.114 \cdot A_V, \tag{4}$$

where $A_V$ is the dust extinction in the visible band, with the factor 0.114 used to convert it into the extinction in the $K_s$ band (J. A. Cardelli et al. 1989). We use a fiducial value of $A_V \sim 0.19$, corresponding to the median value of the sky distribution presented in Y.-K. Chiang (2023). For simplicity of calculations, we assume a null $K$-correction. The values of such a factor in the $K_s$ band and in the redshift range we explore in this work are typically in the [−0.5, 0] range (F. Mannucci et al. 2001), therefore the assumption we make does not significantly affect our results.

The last step is to convert the apparent magnitude in the $K_s$ band into apparent magnitudes in two other bands, namely the mid-IR W1 band and the optical $B$ band. We estimate a correlation between these magnitudes using as input data from the latest version of the Galaxy List for the Advanced Detector Era (GLADE+; G. Dálya et al. 2022). For this, we exclude the sources that have been identified as quasars and the ones estimated to be sites of active star formation.[7] Even if the same galaxy merger that caused the formation of a SMBH binary could also enhance the star formation rate of the host, we make this selection to be self-consistent with the other scaling relations we adopt, which are tailored for ETGs. The apparent magnitudes in the $K_s$ and W1 bands show a strong linear relationship:

$$m_{\rm Ks} = a_3 + b_3 \cdot m_{\rm W1}, \tag{5}$$

where $a_3 = -1.1825 \pm 0.0052$ and $b_3 = 1.0877 \pm 0.0004$ are the best-fit values for the linear fit we perform, and their uncertainties. We also calculate the intrinsic scatter of this correlation as $\sigma_{\rm intr,3} = 0.131$. Similarly, we estimate the relation between $m_{\rm K_s}$ and the apparent magnitude in the $B$ band, $m_{\rm B}$, as

$$m_{\rm Ks} = a_4 + b_4 \cdot m_{\rm B}, \tag{6}$$

where $a_4 = 1.7237 \pm 0.0054$, $b_4 = 0.6517 \pm 0.0003$, and the intrinsic scatter has a value of $\sigma_{\rm intr,4} = 0.431$. As expected, the linear relation between the two IR magnitudes ($m_{\rm K_s}$ and $m_{\rm W1}$) is much tighter than the relation between $m_{\rm K_s}$ and the apparent magnitude in the optical band $m_{\rm B}$, since these two brightnesses correspond to two different, and more distant, regions of the EM spectrum. As further discussed in Section 5, here we choose to estimate empirical relations between different apparent magnitudes based only on observed data. A different possible approach would include using templates of the spectral energy distributions (SEDs) for ETGs and from those estimating the ratios between luminosities in different bands.

### 3.2. Conversion from SMBH Binary Parameters to AGN Apparent Magnitudes

If the SMBH is associated with an AGN, the emission from the AGN will probably outshine the host galaxy. Therefore, converting the SMBH binary parameters into the brightness of the host galaxy in this case requires a different approach with respect to the ETG scenario. As above, we associate the total mass of the binary with the single SMBH mass of the central engine of the AGN. This is a reasonable approximation, since theoretical results have shown that the time-averaged accretion onto a SMBH binary can match that expected for a single SMBH of equivalent mass (D. J. D'Orazio et al. 2013; B. D. Farris et al. 2014; C. Tiede et al. 2025).

[7] In G. Dálya et al. (2022), during the creation of GLADE+, galaxies with active star formation were separated from passive ones using a color cut in the mid-IR band. In particular, a galaxy is considered to be the site of active star formation if W2 − W3 ⩽ 1.5 (T. H. Jarrett et al. 2013; M. E. Cluver et al. 2014).

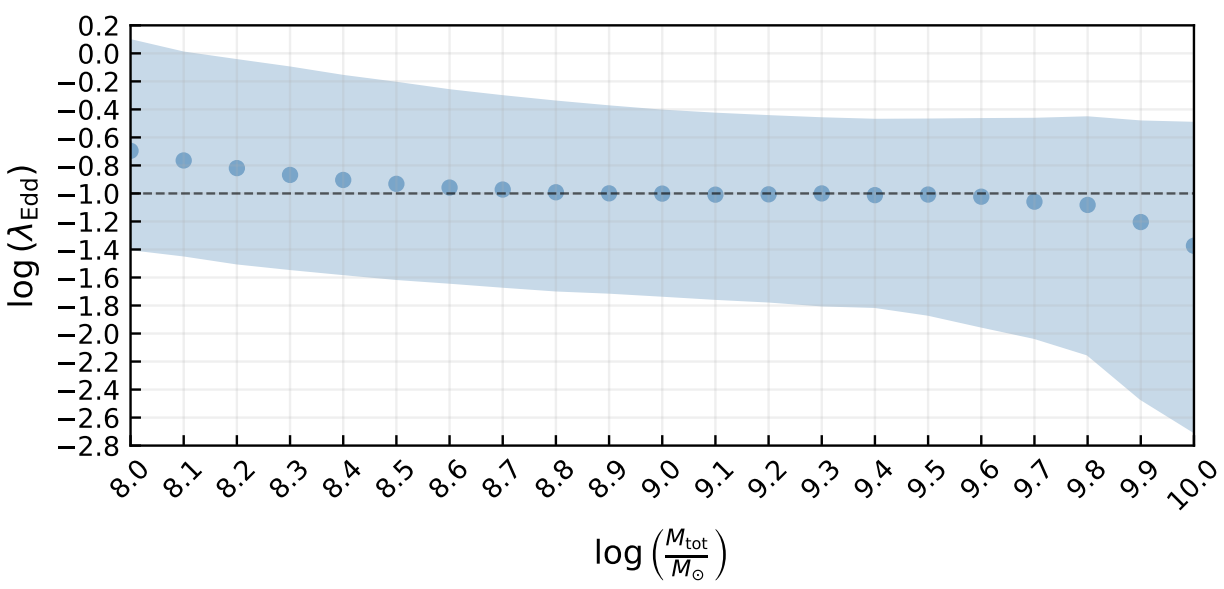

**Figure 3.** Logarithm of the Eddington fraction $\lambda_{\rm Edd}$ as a function of the total mass of the SMBH binary. The round markers indicate the median of the $\lambda_{\rm Edd}$ distribution for each mass bin, while the lower (upper) bound of the shaded area indicates the 5th (95th) percentile. The horizontal dashed line marks the $q = 0.1$ level. The data were taken from the AGN catalog of Q. Wu & Y. Shen (2022).

The first step is to convert the total binary mass into an estimate of the bolometric luminosity, $L_{\rm bol}$, of the AGN. We do so by first calculating the Eddington luminosity, $L_{\rm Edd}$, from the total mass of the binary:

$$L_{\rm Edd} = \frac{4\pi {\rm GM}_{\rm tot} c}{\kappa} \sim 1.26 \times 10^{38} \frac{M_{\rm tot}}{M_\odot}\ {\rm erg\ s^{-1}}, \tag{7}$$

where $G$ is the gravitational constant, $c$ is the speed of light in vacuum, and $\kappa$ is the ratio between the Thomson cross section for electrons and the mass of a proton. We then multiply $L_{\rm Edd}$ by the Eddington fraction, $\lambda_{\rm Edd}$:

$$L_{\rm bol} = \lambda_{\rm Edd} \cdot L_{\rm Edd}. \tag{8}$$

The values of the Eddington fraction we use in our analysis are sampled from probability distributions of $\lambda_{\rm Edd}$ evaluated in distinct mass bins using KDEs. In particular, following the approach of G. Sato-Polito et al. (2025), we use the catalog of quasar properties presented in Q. Wu & Y. Shen (2022). For this catalog, we select sources with masses in the $[10^8 M_\odot,\ 10^{10} M_\odot]$ range and split them in a log-linear grid of 21 bins, each with a 0.1 dex width. For each mass bin, we separately apply a KDE to estimate the probability distribution of the Eddington fraction $\lambda_{\rm Edd}$. Figure 3 shows the distributions of $\lambda_{\rm Edd}$ for the 21 different mass bins, estimated with data from Q. Wu & Y. Shen (2022). The round markers indicate the median value of $\lambda_{\rm Edd}$ for each bin, while the edges of the shaded region indicate the 5th and 95th percentiles of each distribution. Even though this is shown in the figure as a continuous function of the SMBH mass, we emphasize that the distributions are separately calculated in distinct bins.

To convert the bolometric luminosity of an AGN into the luminosities in specific bands, we apply a bolometric correction, BC:

$$\nu_x L_{\nu_x} = {\rm BC}_x \cdot L_{\rm bol}, \tag{9}$$

where the subscript $x$ indicates the band in which the luminosity is estimated in the reference frame of the source. To estimate the bolometric correction, we start from Z. Shang et al. (2011), which provides a sample of AGN with observed SEDs, $\nu^{\rm obs} F_\nu^{\rm obs}$, where $F_\nu^{\rm obs}$ is the observed flux density and $\nu^{\rm obs}$ the observed frequency. We obtain the bolometric luminosities and redshifts of these AGN from J. C. Runnoe et al. (2012). We note that the redshift of this sample ranges between 0.03 and 1.4, covering the entire range we examine in this study. Next, we calculate the rest-frame luminosity, $L_\nu^{\rm rest}$, as a function of the rest-frame frequency, $\nu^{\rm rest}$, as follows:

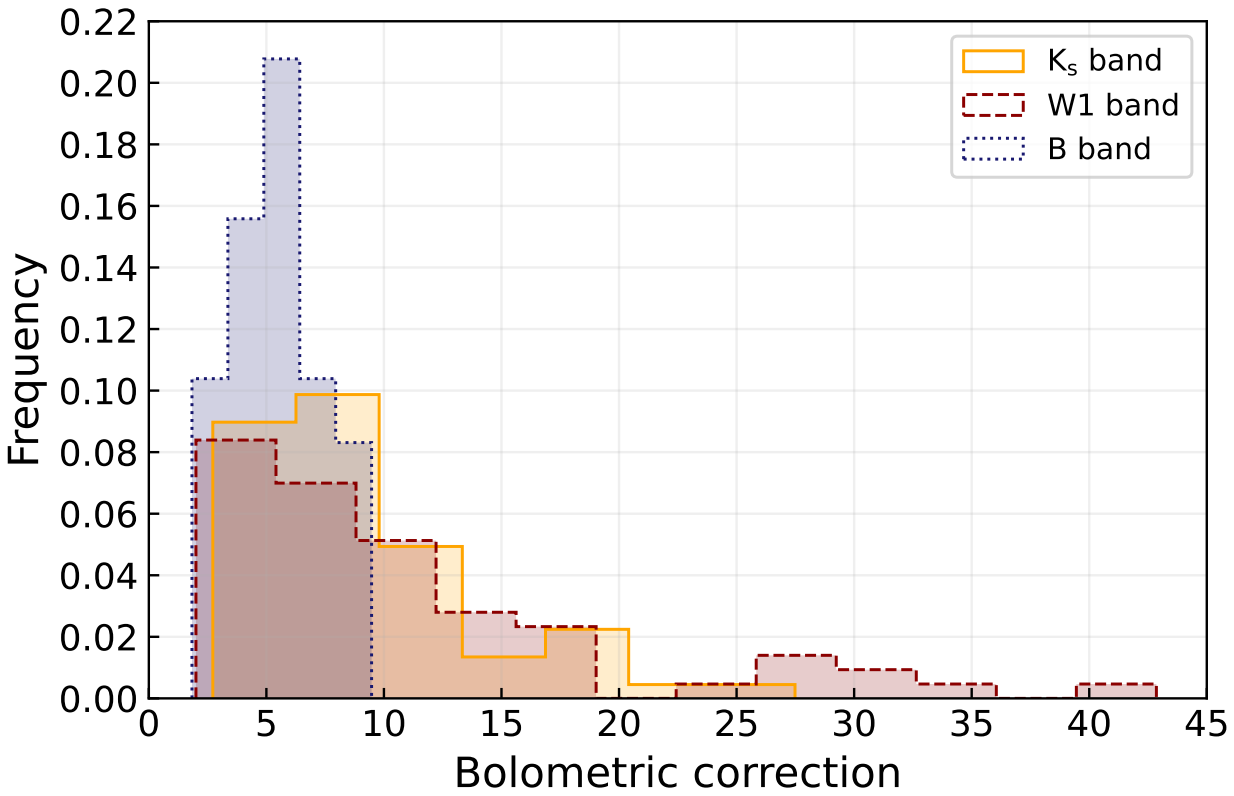

**Figure 4.** Distribution of bolometric corrections for three EM bands, with the yellow solid line corresponding to the 2MASS $K_s$ band, red dashed line to the WISE W1 band, and blue dotted line to the SuperCOSMOS $B$ band. The three histograms are independently normalized. The data used to generate these distributions are taken from Z. Shang et al. (2011) and J. C. Runnoe et al. (2012).

$$\nu^{\rm rest} L_\nu^{\rm rest} = 4\pi d_{\rm L}^2 \nu^{\rm obs} F_\nu^{\rm obs}. \tag{10}$$

We then linearly interpolate $\nu^{\rm rest} L_\nu^{\rm rest}$ as a function of the rest-frame frequency, and evaluate its value for the frequencies that correspond to the fiducial wavelengths of the $K_s$ band (2.159 $\mu$m), the W1 band (3.4 $\mu$m), and the $B$ band (445 nm). We use the bolometric luminosities from J. C. Runnoe et al. (2012) and Equation (9) to obtain the BC in the three bands for each AGN in the sample. The resulting distributions of the estimated BC in each of the three bands are shown in Figure 4. Finally, we approximate these distributions with KDEs, and estimate the luminosities of the host galaxy in each band from their bolometric luminosities and randomly drawing the BC from these distributions.

Having calculated the luminosity of the host in each band, it is straightforward to calculate the absolute magnitude, $M_x$, in the same bands using the following standard relation:

$$M_x = M_{x,\odot} - 2.5 \log\left(\frac{\nu_x L_{\nu_x}}{\nu_x L_{\nu_x,\odot}}\right), \tag{11}$$

where $\nu_x L_{\nu_x,\odot}$ and $M_{x,\odot}$ are the luminosity and the absolute magnitude of the Sun in the $x$ band, respectively. In particular, we use the following values from C. N. A. Willmer (2018): $M_{\rm K\,s,\odot} = 3.27$, $M_{\rm W1,\odot} = 3.26$, and $M_{\rm B,\odot} = 5.44$. The solar luminosities in the three bands are calculated as follows:

$$L_{\nu_x,\odot} = L_{\rm bol,\odot} \times 10^{-0.4(M_{x,\odot} - M_{\rm bol,\odot})}, \tag{12}$$

where $L_{\rm bol,\odot} = 3.83 \times 10^{33}\ {\rm erg\ s^{-1}}$ and $M_{\rm bol,\odot} = 4.74$ are the solar bolometric luminosity and magnitude, respectively.

Finally, we calculate the apparent magnitudes in each band using Equation (4), but with different values for the conversion factor of the extinction. Like in the case of the $K_s$ band, we adopt the values from J. A. Cardelli et al. (1989), and multiply $A_V$ by 0.171 for the W1 band and by 1.337 for the $B$ band.

We emphasize that, while in ETGs the uncertainties in the estimated apparent magnitudes came from the combination of the different uncertainties of the various scaling relations used, in the AGN scenario they primarily arise from the Eddington fraction and the BC. We incorporate these uncertainties into our calculations by randomly sampling values from the $\lambda_{\rm Edd}$ and the BC distributions.

### 3.3. Conversion of a Posterior Function Obtained in Standard GW Analyses into a ($M_{tot}$, $d_L$) Distribution

Typically, a GW search provides estimates on binary parameters such as the chirp mass $\mathcal{M}$, the GW strain amplitude $h_0$, and the frequency of the signal $f_{\rm GW}$. However, these parameters are not the ones that can be used to infer the brightness of the host environment. For this reason, we need to use the joint posterior distribution obtained from the standard GW analysis to transform the measured ($\mathcal{M}$, $h_0$, $f_{\rm GW}$) distribution into a set of ($M_{\rm tot}$, $d_{\rm L}$) combinations.[8] We do so via rejection sampling following the steps here detailed:

1. Calculate the value of $d_L$ that corresponds to each posterior sample of the GW standard analysis as follows:

$$d_{\rm L} = \frac{2({\rm G}\mathcal{M})^{5/3}(\pi f_{\rm GW})^{2/3}}{c^4 h_0}. \quad (13)$$

2. Estimate a continuous 2D probability density function $\mathcal{P}(\mathcal{M}, d_{\rm L})$ from the posterior samples of $\mathcal{M}$ of the GW analysis, and the values of $d_{\rm L}$ calculated in the previous step. From the posterior samples, we obtain a discrete distribution, but through linear 2D interpolation we obtain a continuous function defined in every point of the ($\mathcal{M}$, $d_{\rm L}$) parameter space. This is the target distribution of the rejection sampling and is necessary for the steps below.
3. Extract a random value of $f_{\rm GW}$, $h_0$, $q$, and $M_{\rm tot}$. These random values are individually extracted from the priors used in the GW analysis. Since in this work we use a detection simulation from P. Petrov et al. (2024) as our test case, we adopt the priors from that analysis as listed in their Table 1.
4. From the random draws from the priors, calculate $d_{\rm L,rand}$ from Equation (13) and $\mathcal{M}_{\rm rand}$ using the following equation for the chirp mass:

$$\mathcal{M} = \frac{q}{1 + q^2} M_{\rm tot}. \quad (14)$$

5. Extract a random number, $N_{\rm rand}$, between 0 and the maximum value of the 2D probability density function $\mathcal{P}(\mathcal{M}, d_{\rm L})$.
6. Reject the set of values of $M_{\rm tot}$ and $d_{\rm L,rand}$ obtained in the previous steps from randomly sampling the prior distributions if the random extracted number is higher than the value of the probability density function evaluated at the corresponding set of ($\mathcal{M}$, $d_{\rm L}$), i.e., if $N_{\rm rand} > \mathcal{P}(\mathcal{M}_{\rm rand}, d_{\rm L,rand})$, otherwise accept them.
7. Repeat until the number of accepted samples is equal to the sample size of the target distribution.

The result of this is a set of ($M_{\rm tot}$, $d_{\rm L}$) combinations which reflects the original posterior distribution.

[8] For simplicity of calculations, we choose not to take into consideration other measured binary parameters and the correlations that might exist between them. However, the rejection sampling described in this section can be performed using any amount of parameters, following the same process we detail.

## 4. Results

The goal of this work is to associate the binary parameters obtained from a PTA search (like the chirp mass, the GW frequency, and the strain amplitude) with the brightness of the host galaxy in different bands of the EM spectrum. We first investigate the binary parameter space that is accessible to each PTA configuration and compare it with the reach of EM surveys.

On the EM side, we focus primarily on the optical, near-IR, and mid-IR properties of the hosts, due to the availability in such bands of all-sky surveys with a high level of completeness. In each of these bands we consider a nominal survey which has publicly available galaxy catalogs. In particular, we consider the Two Micron All Sky Survey (2MASS; R. M. Cutri et al. 2003) in the near-IR, the Wide-field Infrared Survey Explorer (WISE; E. L. Wright et al. 2010) in the mid-IR, and the SuperCOSMOS Sky Survey (N. C. Hambly et al. 2001) in the optical band. The possibility of using different surveys is discussed in Section 5, together with their potential advantages and drawbacks.

Next, we show a practical application of the method we developed. We show how these conversions can be used in real host-galaxy searches, applying them on one of the mock PTA detections detailed in P. Petrov et al. (2024). We obtain realistic probability distributions of the apparent magnitudes of the host galaxy in the three EM bands we investigate.

### 4.1. Comparison IPTA Reach with EM Surveys

Using the three different IPTA configurations (IPTA_20, IPTA_25, and IPTA_30) described in Section 2.1 and the method to calculate the S/N of an injected SMBH binary detailed in Section 2.2, we calculate the maximum luminosity distance from Earth at which a SMBH binary can be detected with a $S/N \geqslant 8$ as a function of its total mass. As mentioned in Section 2.3, we simulate binaries with circular orbits and face-on orientations, and a fiducial GW frequency of $10^{-8}$ Hz.

To estimate the range of possibilities for the detectable binary parameter space for the three different baselines and PTA configurations, we inject binaries in the best and worst sky location of each PTA, also varying the mass ratio from an optimistic scenario of equal-mass binaries with $q = 1$ to a more pessimistic case with a mass ratio of $q = 0.1$. The results of these calculations are shown by the black and the gray lines in Figures 5 and 6. In particular, the black lines show the reach of the IPTA_20 data set considering different sky positions and mass ratios for the source, with the solid line for the best-case scenario, i.e., best sky location and $q = 1$, dotted for the worst-case scenario, i.e., worst sky location with $q = 0.1$, and, for the intermediate cases, dashed and dashed–dotted lines for the best sky position with $q = 0.1$ and worst sky position with $q = 1$, respectively. The gray dashed–dotted and doted lines show the maximum reach of IPTA_25 and IPTA_30, respectively, in the best-case scenario. The gray shaded area marks the region of the parameter space that can be probed with IPTA_20 under the most optimistic assumptions. We emphasize that here we focus on covering the entire range of detectable binaries, without considering the most likely properties of the first detection (see, e.g., B. Bécsy et al. 2022; E. C. Gardiner et al. 2025; R. J. Truant et al. 2025).

We find that the smallest mass that is detectable by IPTA_20 for luminosity distances greater than 10 Mpc is $10^{8.6} M_\odot$. The maximum distance at which a binary with total mass $10^{10} M_{\odot}$ (the maximum we consider; see Section 2.3) can be detected is $10^{3.3}$ Mpc. The difference in the maximum distance (for a fixed binary mass) between the best and worst sky location is approximately 0.3 dex. Moreover, an equal-mass binary can be detected from a luminosity distance which is approximately 0.5 dex greater than the $q = 0.1$ case (for fixed total binary mass and sky position). The increased baselines and the addition of new pulsars lead to a milder increase in the maximum probed distance. From IPTA_20 to IPTA_25, as well as from IPTA_25 to IPTA_30, we observe a similar increase of approximately 0.1 dex (considering the best sky location and equal-mass binaries).

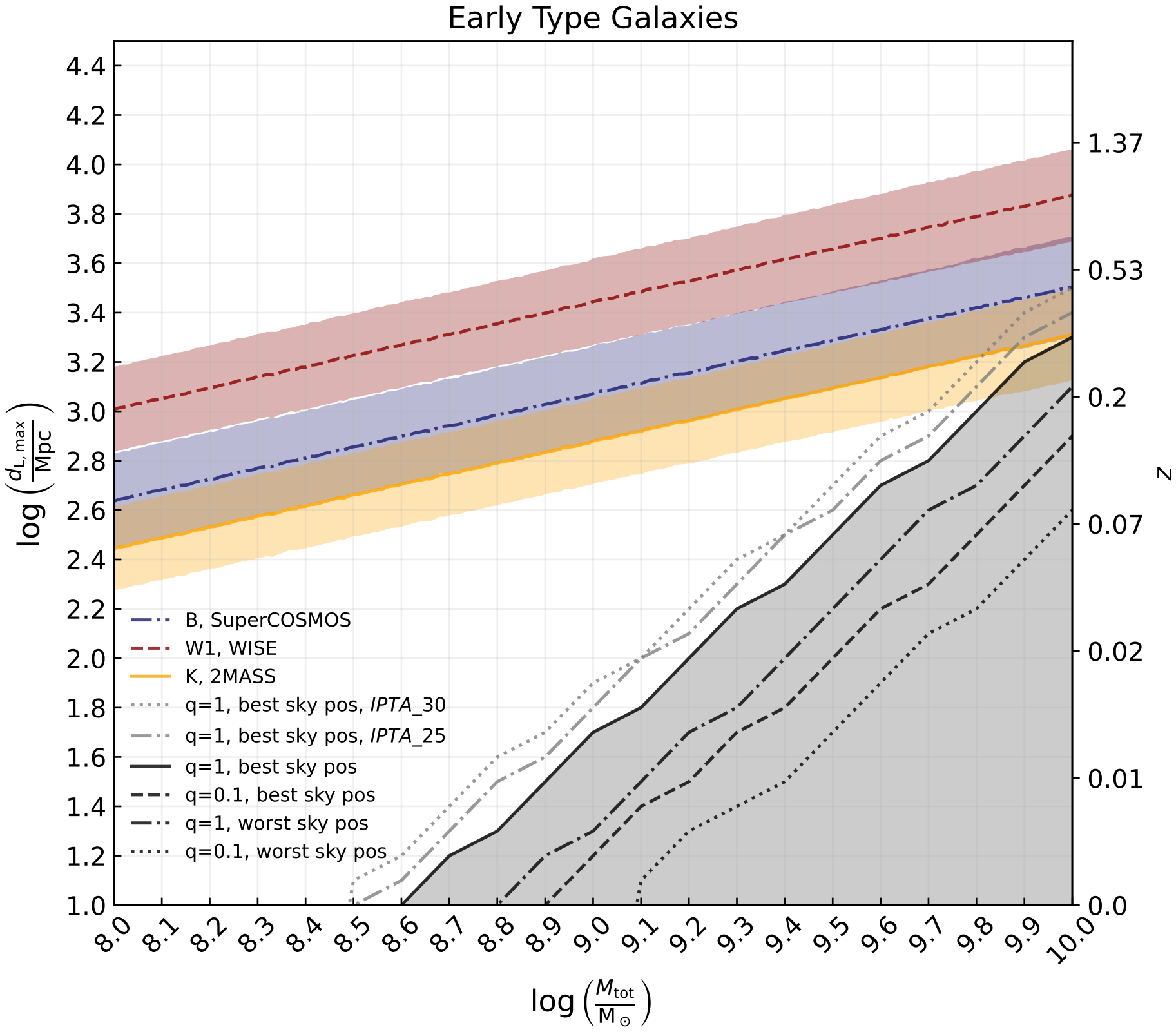

**Figure 5.** Comparison of the maximum distance at which PTAs can detect individual binaries as a function of their total mass with the maximum distance at which EM surveys could detect their host galaxies if they reside in ETGs. The gray shaded region delineates the range of parameters IPTA_20 can probe under the most optimistic circumstances (best sky location and equal-mass binaries). The other black lines show the same parameter space for IPTA_20, but for less optimistic assumptions, i.e., best sky position with unequal-mass binaries (dashed line), worst sky position with equal-mass binaries (dashed–dotted line), and worst sky position with unequal-mass binaries (dotted line). Similarly, the gray lines denote the accessible binary parameter space for future IPTA configurations, with dashed–dotted (dotted) gray line for IPTA_25 (IPTA_30), respectively, under the most optimistic scenario for detection. The color lines demonstrate how far different all-sky EM surveys can detect a regular ETG as a function of the mass of the SMBH binary it could host, with yellow solid line for 2MASS, blue dashed–dotted line for SuperCOSMOS, and dark red dashed lines for WISE.

The maximum distance for IPTA_30 corresponds to a redshift of $z \sim 0.5$ at the maximum total mass of $10^{10} M_{\odot}$, in agreement with theoretical expectations (P. A. Rosado et al. 2015; L. Z. Kelley et al. 2018). This result depends both on the PTA configuration we employ and the S/N threshold we choose. However, given that the gain in sensitivity is not dramatic between IPTA_20 and IPTA_25 (or between IPTA_25 and IPTA_30) with the addition of 5 yr of data and 42 pulsars, the details of the PTA construction are unlikely to be very significant. Had we chosen a lower S/N threshold (e.g., $S/N \sim 5$), we would have been able to probe binaries at higher distances. However, the PTAs we use were built to represent realistic PTAs in the near and intermediate future, and the choice of $S/N = 8$ is motivated by previous work as the threshold above which we should be able to localize sources. We also note that, even with a more sensitive PTA configuration, consisting of 200 pulsars with 30 yr of Square Kilometre Array data and a lower S/N threshold, R. J. Truant et al. (2025) found that most of the resolved binaries lie at relatively low redshifts ($z < 0.5$), but reaching a maximum redshift up to $z \sim 2$.

In Figure 5, we show the maximum luminosity distance at which a regular ETG is bright enough to be observed by 2MASS, WISE, or SuperCOSMOS as a function of the mass of the SMBH binary it might host. These distance limits for the EM surveys have been calculated using the steps described in Section 3.1, assuming the following values for limiting apparent magnitudes: $m_{K_s} = 14.3$ for 2MASS, $m_{W1} = 16.83$ for WISE, and $m_B = 20.79$ for SuperCOSMOS. The chosen magnitude limit for 2MASS is the one that achieves a detection with $S/N \geqslant 10$ for unconfused point sources outside the Galactic plane, i.e., $|b| > 10°$ (M. F. Skrutskie et al. 2006). Similarly, for WISE the limiting magnitude is calculated from comparisons with external photometry from 2MASS and Spitzer (M. W. Werner et al. 2004), and corresponds to a detection with $S/N \geqslant 5$ (E. L. Wright et al. 2010).[9] The chosen threshold for SuperCOSMOS corresponds to a detection significance of $5\sigma$ in sky regions far from the Galactic plane, i.e., $|b| > 60^\circ$ (J. A. Peacock et al. 2016). We further discuss the magnitude cuts and the homogeneity of surveys in Section 5. The solid, dashed, and dashed–dotted colored lines in the figure mark the mean value of the maximum distance at which SuperCOSMOS, 2MASS, and WISE can detect galaxies with a certain SMBH mass (or, equivalently, a certain total binary mass), respectively. The width of the colored bands represents the standard deviation of the distribution obtained by incorporating the uncertainties in the conversion from binary total mass to apparent magnitude of the galaxy detailed in Section 3.1.

[9] We choose to use the conservative value for the magnitude limit of the original WISE release; future works might use the catalog of unWISE (D. Lang 2014; E. F. Schlafly et al. 2019), constructed starting from the WISE original data, coadding all the publicly available WISE imaging in the 3–5 $\mu$m band.

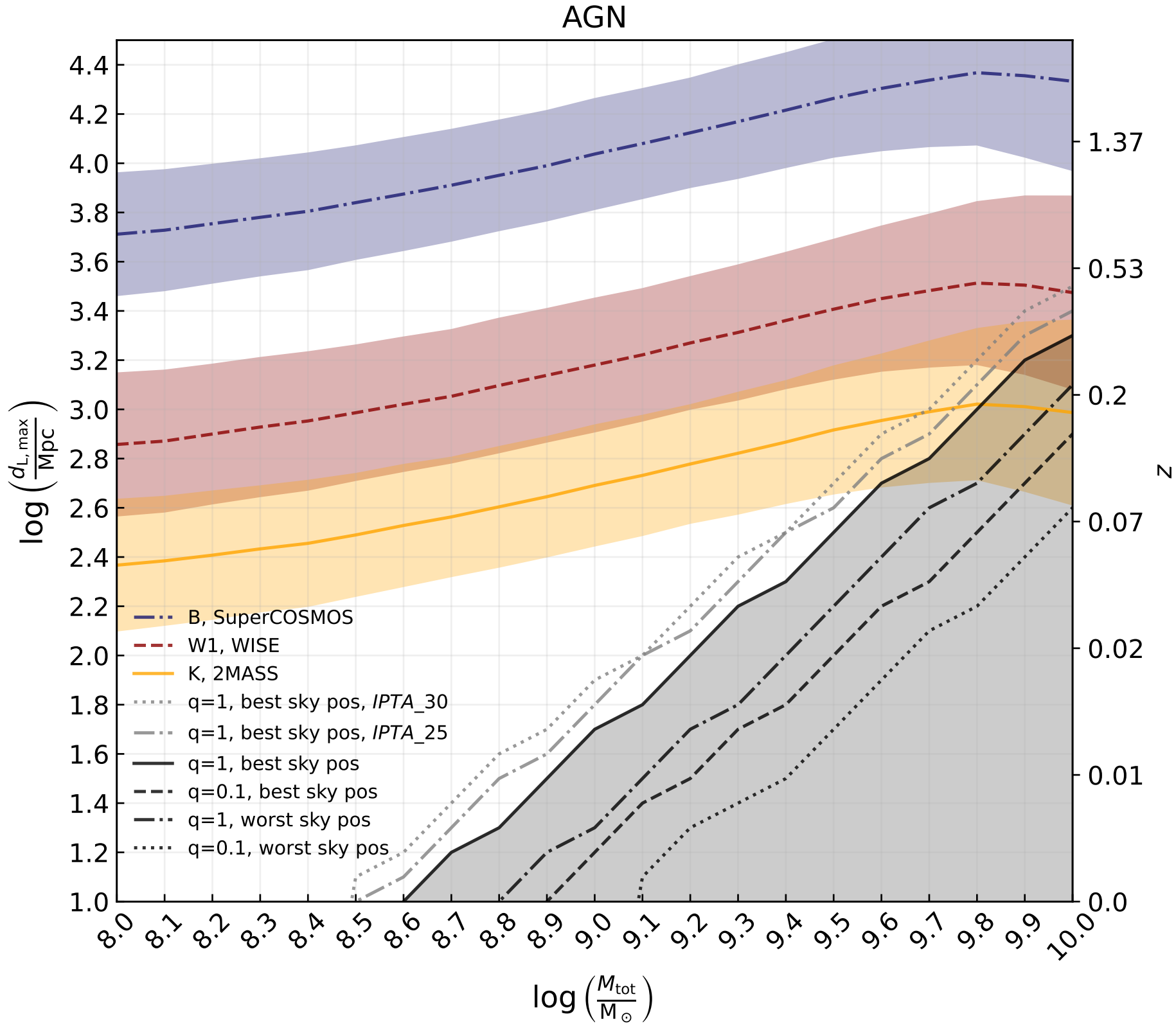

**Figure 6.** Comparison of the maximum distance PTAs can detect binaries vs. how far EM surveys can detect the hosts of these binaries if they reside in AGN as a function of binary total mass. The color-coding and line styles follow that of Figure 5.

Figure 6 shows the same calculations as Figure 5, but for the scenario in which the SMBH binary is hosted in a galaxy with an AGN. We adopt the same magnitude limits for the EM surveys, and the color bands represent the uncertainty in the conversion from binary total mass to AGN brightness, as described in Section 3.2. This uncertainty arises from uncertainties in the Eddington fraction and the BC, and is overall larger compared to the ETG case. Moreover, we observe a slight turnover in the maximum distance EM surveys can cover for the higher masses, mainly driven by a similar turnover in the Eddington fraction distributions (see Figure 3).

By comparing Figures 5 and 6, we see that, for binaries of comparable mass, AGN and ETGs have very similar near-IR magnitudes. This may indicate that when searching for the host galaxy, it may be advantageous to use near-IR magnitudes, since the conversion from the binary properties to EM observables is more weakly dependent on the type of the host. On the other hand, AGN have significantly higher optical and lower mid-IR brightnesses compared to ETGs. Typically, mid-IR emission in AGN comes from the hot dusty torus which surrounds the SMBH, whereas in ETGs the mid-IR emission is dominated by the old stellar population (M. S. Clemens et al. 2009; L. Burtscher et al. 2015). As a result, the mid-IR magnitudes may be dominated by stellar emission, even if a relatively low-luminosity AGN is active in the center of the galaxy. However, when considering Figures 5 or 6 separately, the differences between the reaches in the various EM bands are mainly caused both by how the total luminosity is distributed among the different bands and by the different threshold magnitudes of the specific surveys we consider in our calculations. For example, the fact that WISE performs better than SuperCOSMOS in the ETG scenario does not imply that any past or future mid-IR survey is expected to perform better than any optical survey. The results shown in these figures are therefore not to be interpreted as an indication on what type of host is the preferred one, but rather on the completeness of the different EM surveys in the region of the parameter space that can be explored by PTAs.

We find that WISE and SuperCOSMOS are expected to detect all the potential host environments of binaries within the sensitivity volume of IPTA_20, regardless of the type of such hosts. As far as 2MASS is concerned, and for the IPTA_20 sensitivity volume, the survey starts to become incomplete for hosts of binaries with a total mass greater than $10^{9.8} M_\odot$ and a luminosity distance from Earth greater than 1 Gpc. However, as the sensitivity of PTAs increases in IPTA_25 and IPTA_30, allowing us to measure GW signals from greater distances, the incompleteness of the surveys to binary hosts will become more important.

### 4.2. *Estimation of Host Magnitude Distributions for a Simulated IPTA Detection*

In the previous section, we calculated the range of binary parameters PTAs can probe in the near and intermediate future and examined whether EM surveys can cover their host galaxies. Now we present a practical application of this method. Specifically, we demonstrate how the posterior distributions of SMBH binary parameters inferred by a PTA search can be converted into distributions of the brightness of the host galaxy. For this, we use the posterior samples of one of the simulated detections presented in P. Petrov et al. (2024).

In particular, we use the posteriors of a binary in the galaxy J19231198–2709494, and detected with $S/N = 8$. This galaxy is located at a luminosity distance from Earth of 276.6 Mpc, and the simulated binary has a total mass of $10^{9.53} M_{\odot}$. We choose this particular simulated GW signal since it employs the same S/N threshold we use throughout this work, which also represents the more realistic scenario for the first real detection (as opposed to the optimistic cases with $S/N = 15$ also presented in P. Petrov et al. 2024). In addition, this is one of the five injections with $S/N = 8$ in which the localization area is well constrained. We do not expect that choosing a different GW injection would produce qualitatively different results, as long as the GW signal is well localized. However, had we chosen one of the high-S/N injections, we expect that it would produce significantly narrower distributions of the apparent magnitudes of the host environment.

We use the rejection sampling described in Section 3.3 to create a 2D posterior distribution of the binary total mass and luminosity distance ($M_{\rm tot}$, $d_{\rm L}$). For each sample of this distribution, we use the methods described in Sections 3.1 and 3.2 to estimate the apparent magnitudes in the $K_s$, W1, and $B$ bands. To take into account all the uncertainties involved in turning binary properties into brightnesses of the host environment, for each sample we estimate 1000 different values of the magnitudes drawing from distributions that incorporate the uncertainties in each step of the conversion.

The outcome of this process is presented in Figure 7. The solid (dashed) lines in the contour plots, as well as in the 1D marginalized distributions, show the results obtained assuming the host environment is a regular ETG (AGN). The thin vertical lines in the histograms and the plotted levels of the contours correspond to the 5th, 50th, and 95th percentiles of the corresponding distributions. The values of the median and its distance from the 5th and 95th percentiles of the marginalized distributions are also reported above each panel, for the case of AGN hosts. The wider colored vertical lines in the 1D histograms mark the limiting magnitude of the three different EM surveys used in this work, using the same color and line style coding as in Figures 5 and 6. The marginal posterior distribution in the 2D space parameterized by the luminosity distance and the total mass is obtained solely from the mock GW detection and analysis, and does not depend on any of the calculations done to estimate the brightness of the host galaxy. On the contrary, the width of the posterior distributions of the apparent magnitudes depends on how precisely the binary parameters are estimated in the standard GW analysis.

As expected, for this particular example the majority of these distributions (with the exception of the tails roughly at the percent level) are above the detection thresholds of the 2MASS, WISE, and SuperCOSMOS surveys, which means that the majority of potential hosts would be present in the all-sky catalogs of these surveys unless they fall close to the Galactic plane or are obscured. We note that a significant advantage of the IR bands is that they are typically unaffected by obscuration caused by the dusty torus.

The conversion from binary parameters to expected photometric properties of the host leads to marginally different distributions if the galaxy is a regular ETG compared to an AGN. Likewise, Figures 5 and 6 show that these two assumptions lead to different expected maximum luminosity distances at which the EM counterpart can be detected as a function of the observed band. Given the estimated sensitivity of IPTA data sets in the next 10 yr, which could detect binaries with total masses of $10^{10} M_{\odot}$ up to redshifts $z \sim 0.5$, the surveys we consider are mostly complete for both types of potential hosts. However, these different estimates are not suited for assessing whether one of these two scenarios is physically preferred over the other. A practical application of the method presented in this work will likely involve one catalog of ETGs and one catalog of AGN. The differences in the distributions shown in Figure 7 will result in different selections or ranking criteria for the identification of the true host, after the sky map of the GW detection has been cross-matched with such galaxy catalogs. Such ranking processes will be based on the posterior distributions of the apparent magnitudes and are meant to be done in parallel for the two different host types we take into account, without, as mentioned before, assessing which of the two is favored over the other.

To test how efficient the presented method is when applied to GW detections and cross-matched with galaxy catalogs, we perform a simple yet representative calculation. Among the posterior distributions of apparent magnitudes shown in Figure 7, we consider the broadest, which corresponds to the $B$-band magnitude $m_{\rm B}$ in the ETG scenario. The distance between the 5th and the 95th percentiles of such a distribution is $\sim$16.8 mag, while for the other posterior distributions this difference is between $\sim$10.0 and $\sim$12.0 mag. We compare the 95th percentile of the distribution of $m_{\rm B}$ in the ETG scenario with the distribution of the apparent $B$-band magnitude from the WISE x SuperCOSMOS Photometric Redshift Catalog (M. Bilicki et al. 2016), which contains a total of 20,416,142 galaxies and is the result of a cross-match between two of the surveys we consider in this study, namely WISE and SuperCOSMOS. We find that more than $\sim$84% of the galaxies have observed magnitudes above the 95th percentile threshold. Therefore, $\sim$84% of the galaxies within the localization area of the GW detection could be rejected as potential hosts, since they would be dimmer than expected. This estimate is obtained assuming that the distribution of apparent magnitudes of the galaxies within the GW localization area reflects the overall distribution of the catalog. This is a reasonable assumption, given the expected large number of galaxies within such a localization area.

## 5. Discussion

The method presented in this work aims to facilitate the identification of host galaxies of SMBH binaries that may be detectable by PTAs in the near and intermediate future. This method enables us to select potential host galaxies based not only on their sky position and luminosity distance from Earth, but also on their brightness in different regions of the EM spectrum. Here we discuss the caveats of our analysis, which will need to be addressed in future studies for such a selection to be more effective.

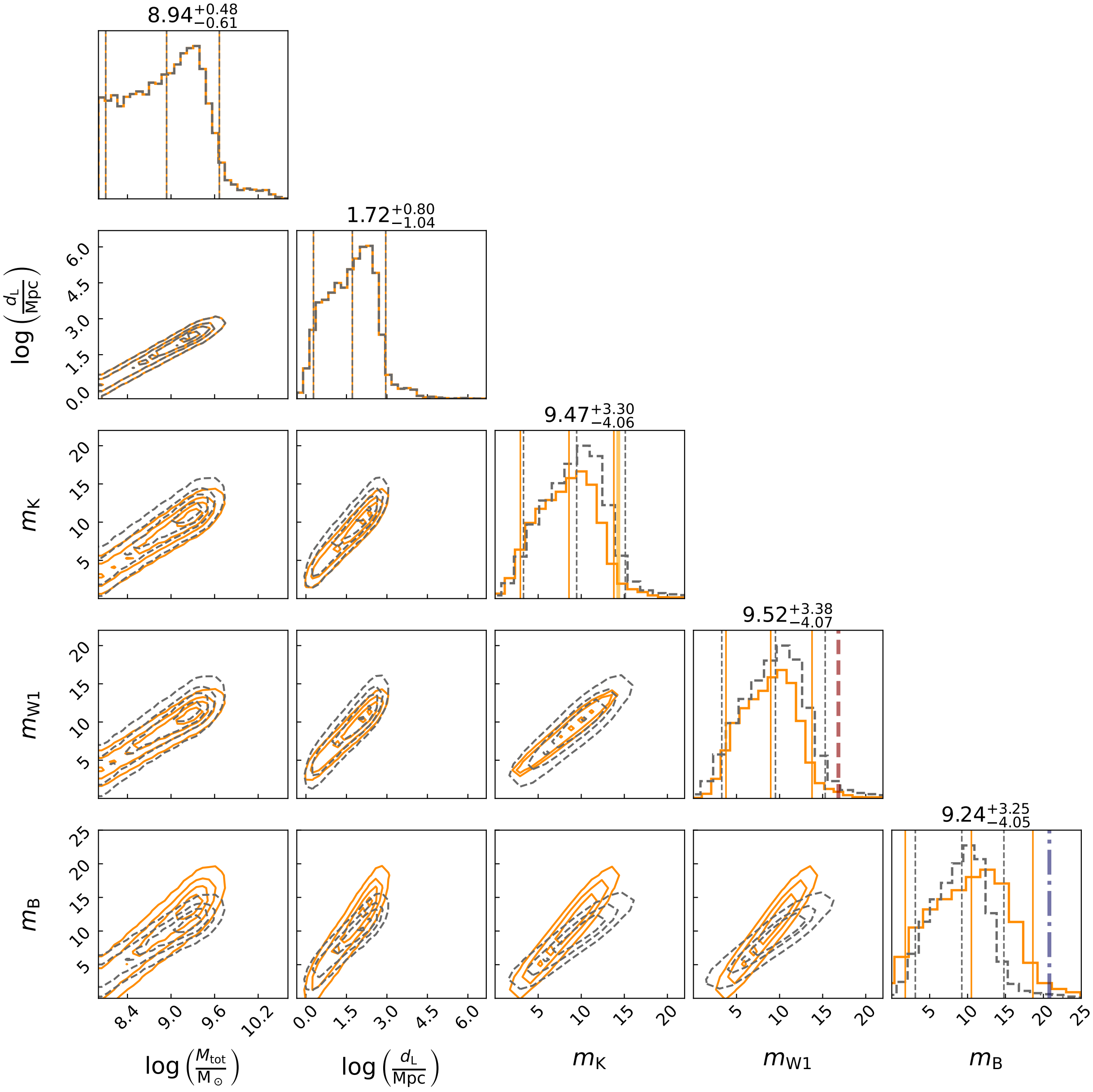

**Figure 7.** Distributions of binary total mass, luminosity distance from Earth, and host-galaxy apparent magnitudes in the $K_s$, W1, and $B$ bands for a mock GW detection of a SMBH binary hosted in the galaxy J19231198–2709494. The GW analysis was performed by P. Petrov et al. (2024), from which we obtain the joint ($M_{tot}$, $d_L$) distribution via rejection sampling, while the apparent magnitude distributions are calculated via the methods presented in Sections 3.1 and 3.2. The orange solid lines represent the scenario of the host being a regular ETG, while the dashed gray line corresponds to the AGN host scenario. Contours show the 5th, 50th, and 95th percentiles of the 2D distributions. The same percentiles are shown with the thin vertical lines in the marginalized distributions, with the numerical values at the top corresponding to the AGN host scenario. The vertical yellow, red, and blue lines in the marginalized distributions mark the nominal magnitude limits we use in this work for 2MASS, WISE, and SuperCOSMOS, respectively, and use the same line style coding as in Figures 5 and 6.

*Survey depth.* We have assumed nominal magnitude limits for the EM surveys we consider in this study. As implied even from the definition of limiting magnitudes presented in Section 4.1, the depth of an all-sky EM survey is not entirely isotropic. For instance, fields that are observed more often typically achieve greater sensitivity. Therefore, depending on the observing strategy of the survey, the limiting magnitude may depend on the sky position, but this effect is typically not very significant.

Of greater importance is the fact that most all-sky EM surveys either completely avoid the Galactic plane or apply selection masks to remove sources from that region in their data releases due to the high level of contamination from stars. This is significant, because the sensitivity of PTAs is better in regions close to the Galactic plane, where most of the better monitored pulsars are clustered; in all three IPTA configurations we consider, the most sensitive sky position is close to the plane of the Milky Way. As a result, in that sky location PTAs are able to probe a bigger volume, thus increasing the chances for a detection. This may present a significant limitation in identifying the host galaxy, given the low level of completeness of EM surveys in that area, combined with the fact that this is typically a sizeable fraction of the entire sky ($\sim$17% if one assumes the Galactic plane to be the region within 10° from the Galactic equator, i.e., $|b| \leqslant 10°$, where $b$ is the Galactic latitude).

In a future study, we plan to investigate the relevance of this effect in the search for host galaxies of PTA sources. For this,

it is necessary to include a realistic PTA configuration like the one we simulated, and instead of exploring the entire range of possibilities, to focus on assessing the most probable parameters of the first few detectable binaries (including their sky location) and on estimating the number density of potential hosts within the localization area. This can be achieved with simulations of realistic SMBH binary populations and the use of available galaxy catalogs. We expect that the completeness of different catalogs with respect to the 3D localization volume of the GW signal will strongly depend on the sky position of the binary. Taking into consideration realistic PTA detections and cross-matching their sky maps with galaxy catalogs with measurements of the completeness as a function of luminosity distance and sky position will also allow an estimate of how much information is added when, in the search for the host galaxy, probability distributions of the brightness in different regions of the EM spectrum are taken into account. In particular, it will be possible to assess what fraction of the potential hosts that are selected based only on their position will be rejected during the photometry-based selection.

*SMBH–galaxy correlations.* One important component of the conversion from binary properties to observables of the host galaxy is the scaling relation used to translate the SMBH mass into the bulge mass (i.e., the $M_{\rm BH}$–$M_{\rm bulge}$ correlation). As mentioned in Section 3.1, we use the best-fit parameters of the correlation presented in N. J. McConnell & C.-P. Ma (2013). However, a wide range of values is reported in the literature, depending on the data set used to derive the correlation (see, e.g., N. Häring & H.-W. Rix 2004; J. E. Greene et al. 2010; Z. Schutte et al. 2019). We also note that the recently discovered GW background has a higher amplitude than previously expected, which may suggest that galaxies have more massive SMBHs than previously thought based on those correlations (G. Agazie et al. 2023c; EPTA Collaboration et al. 2023). This trend would make our predictions about the maximum distance more pessimistic (since the same mass would correspond to a smaller, less luminous host). However, it is expected that the inference on these correlations will improve as we improve the constraints on the GW background.

The intrinsic scatter in the $M_{\rm BH}$–$M_{\rm bulge}$ relation is the main source of uncertainty in the estimation of the maximum distance covered by the different EM surveys, contributing approximately 93%, 92%, and 83% of the total uncertainty one obtains when converting a fixed value of the binary total mass into the reach of 2MASS, WISE, and SuperCOSMOS, respectively. Such total uncertainty is represented by the width of the yellow, red, and blue shaded regions in Figure 5. However, repeating the analysis described in Section 4.2, the results of which are shown in Figure 7, without taking into consideration any uncertainty (nor intrinsic scatter) in the $M_{\rm BH}$–$M_{\rm bulge}$ relation, we find that the standard deviation of the resulting distribution of apparent magnitudes is reduced only by $\sim$2%. Therefore, we conclude that the main source of uncertainty in the estimated apparent magnitudes comes from the width of the posterior distributions of the binary parameters from the standard GW analysis.

*Inter-band correlations and SEDs.* In the case of ETG galaxies, we create an empirical correlation between the apparent magnitude in the $K_s$ band, which we estimate from the stellar mass, and the apparent magnitude in the other two bands using observed data from the GLADE+ galaxy catalog. When propagating the uncertainties from each step of the calculation, we find that the uncertainties associated with inter-band magnitude conversion contribute only marginally to the total uncertainty of the maximum distance we can probe with EM surveys. A more robust approach could be to use synthetic SED templates estimated from large ensembles of ETGs and use those to calibrate the luminosities in the W1 and $B$ bands starting from the $K_s$-band luminosity. Overall, we expect this to have a negligible effect on the final results.

*Redshift.* In this work, we have neglected effects introduced by the redshift evolution of various quantities, but we expect that none of these can significantly alter our conclusions. First, in the GW signal, we measure redshifted quantities like the chirp mass $\mathcal{M}^{\rm obs} = (1+z)\mathcal{M}^{\rm rest}$ and the GW frequency $f_{\rm GW}^{\rm obs} = f_{\rm GW}^{\rm rest}/(1+z)$. However, PTA searches typically ignore the redshift (i.e., they set it to zero), because current sensitivity is limited to the relatively nearby Universe. For instance, the luminosity distance of the source examined in Section 4.2 corresponds to a low redshift value ($z \sim 0.06$). We note that even in the most optimistic scenario in IPTA_30, the sources have moderate redshifts ($z < 0.5$), therefore the effect is relatively small. However, as the reach of PTAs increases, this redshift dependence should be taken into account in the future.

In addition, when constructing the empirical distributions of the Eddington ratios and of the BC with data taken from Q. Wu & Y. Shen (2022), Z. Shang et al. (2011), and J. C. Runnoe et al. (2012), we do not take into account the possible redshift dependence of these parameters. To ensure a large enough sample, we use the complete data sets from the original publications. Even though we did not perform any redshift cuts, we do not expect this to introduce any significant bias in our results, because the redshift evolution of these parameters is weak (see Figure 6 of G. Sato-Polito et al. 2025). A redshift dependence is also expected to exist in the best-fit values of the parameters of the $M_{\rm BH}$–$M_{\rm bulge}$ correlation. We do not take this into account in this work, but again, since our analysis is restricted to a relatively small redshift range in the local Universe, within which the best-fit values of the scaling relations are not expected to vary noticeably (M. M. Kozhikkal et al. 2024), our results will likely be unaffected by this choice. The last redshift dependence we do not include is the $K$-correction factor in the relation between absolute and apparent magnitudes of the galaxies. As explained in Section 3.1, this choice does not significantly affect our results. However, a correct characterization and inclusion of this factor will be present in future works that aim to cross-match sky maps of mock GW detections of SMBH binaries and estimated distributions of the apparent magnitudes of their hosts with observed catalogs.

Beyond the caveats presented above, we also present some opportunities for future extensions of the current study.

*X-ray observations.* In this work, we focused on the (near and mid) IR and optical observables of the host galaxies of PTA sources, but a similar analysis can be extended to other regions of the EM spectrum, such as the X-ray. However, calculating the X-ray brightness of the host galaxy comes with significant challenges. For instance, the correlation between the X-ray luminosity and other properties of the host environment, like its stellar mass or its absolute magnitude in other bands, is poorly constrained. This is true for both ETGs (S. C. Ellis & E. O'Sullivan 2006; Y. Zhang et al. 2024) and AGN, because of the big spread in the bolometric correction distribution (J. C. Runnoe et al. 2012). Given the sensitivity of X-ray telescopes like Chandra (M. C. Weisskopf et al. 2000), it is likely that these massive galaxies at such low redshifts can be detected (whether they host an AGN or not). However, a systematic search of the host galaxy in an X-ray catalog is unlikely to be fruitful; the aforementioned uncertainty in converting the binary properties into X-ray fluxes would make any host association extremely tentative. Even though X-rays may not be optimal for the initial selection of candidate host galaxies, follow-up observations to uncover X-ray binary signatures may offer a promising route, once the sample is limited to a manageable number of potential hosts.

*Radio observations.* For reasons similar to the above, we did not consider radio observations either, even though high-resolution observations in this band could offer the definitive confirmation of a host galaxy. If both components of the binary system are bright enough in the radio band to be detectable, very-long-baseline interferometry could resolve the binary, providing the "smoking gun" signature. For example, the angular separation of a SMBH binary with a total mass of $10^{9.5}M_{\odot}$, orbiting at a frequency of $10^{-8}$ Hz, and located at a luminosity distance of $10^{2.5}$ Mpc (the distance at which we estimate such a system to be detectable by the IPTA_20 configuration with $S/N = 8$, if it is located in the best possible sky position and has a mass ratio of $q = 1$) is approximately 20 $\mu$as, which is slightly bigger than the angular resolution of the Event Horizon Telescope: 19 $\mu$as at 345 GHz (Event Horizon Telescope Collaboration et al. 2019). Therefore, depending on the binary parameters, the radio band could offer excellent opportunities for follow-up observations. However, it is important to note that a scenario in which a binary is directly imaged in the radio band requires both components to be radio-loud, and since most AGN are radio-quiet, such binaries are expected to represent only a small fraction of the overall population.

*Additional EM surveys.* In this study, we considered three all-sky EM surveys. We demonstrated that these surveys are able to cover most, but not all, of the massive hosts of PTA sources within the volume PTAs can probe. For IPTA_25 or IPTA_30, which can detect binaries up to redshift $z \sim 0.5$, these EM surveys will become increasingly more incomplete. However, upcoming telescopes, like the Vera C. Rubin Observatory (Ž. Ivezić et al. 2019) in the optical and the planned Nancy Grace Roman Space Telescope (R. Akeson et al. 2019) in the near-IR, are characterized by considerably higher limiting magnitudes, and will therefore be able to observe PTA host galaxies to even greater distances with respect to the surveys we took into consideration. Galaxy catalogs from other telescopes, like the Spectro-Photometer for the History of the Universe, Epoch of Reionization and Ices Explorer (or SPHEREx; B. P. Crill et al. 2020), Euclid (Euclid Collaboration et al. 2025), and the Dark Energy Spectroscopic Instrument (DESI Collaboration et al. 2016), are also expected to play a significant role in the search for the host environments of PTA sources.

## 6. Summary and Conclusion

We presented a method to convert the properties of SMBH binaries detectable by PTAs into the EM properties of their host environments, assuming the hosts to be either regular ETGs or AGN. We designed three PTA configurations that represent realistic scenarios for IPTA data sets in the near and intermediate future, and assessed the binary parameter space that is accessible to each. Then we examined whether three EM surveys, namely 2MASS, WISE, and SuperCOSMOS, are sensitive enough to detect the binaries' host galaxies.

We find that IPTA_20 (which resembles the upcoming data release of the IPTA collaboration) can detect binaries at distances of 2Gpc (under the most optimistic assumptions). The potential host galaxies of all such binaries are above the nominal brightness threshold for WISE and SuperCOSMOS, while 2MASS will be incomplete to host galaxies of binaries with total masses greater than $10^{9.8}M_{\odot}$, which can be detected to a luminosity distance from Earth greater than 1 Gpc. As the sensitivity of PTAs increases, they will be able to probe binaries at larger and larger distances. IPTA_30, with a baseline of 30 yr and 200 pulsars, can reach out to $z \sim 0.5$. At these distances, the EM surveys will become increasingly more incomplete to their hosts in the $K_s$ and $B$ bands in the case of ETGs, and in the $K_s$ and W1 bands in the case of AGN. Figures 5 and 6 summarize the results of the comparison between the maximum distance at which PTAs can detect individual binaries versus how far EM surveys can detect their hosts.

We also present a practical example of how to convert a posterior distribution from a standard GW analysis into predictions of the brightness of the host environment in different wave bands. The results of this example are shown in Figure 7. Such predictions will be crucial when PTAs make a detection of a resolved SMBH binary, allowing us to select the candidate host galaxies within the GW localization area. This selection will rely not only on the positions of the potential hosts but also on their EM properties (e.g., luminosities/magnitudes), which are readily available in galaxy catalogs. The importance of calculating the expected brightness in different bands of the EM spectrum, even if they are found to correlate with each other, lies for example in the fact that different surveys have in general different levels of depth and different sky coverage, especially in proximity to the plane of the Milky Way.

Searches like this, which include information about the photometry of the host galaxy, serve as a starting point for choosing which objects to focus on for follow-up analyses. These will likely include more detailed archival searches and new observational campaigns to look for EM signatures that hint at the presence of a binary.

## Acknowledgments

The authors thank Polina Petrov for providing the posterior distributions of the GW analyses of the mock binary detections in P. Petrov et al. (2024); Alberto Sesana, David Izquierdo-Villalba, and Riccardo Truant for their useful comments; and the referee for suggesting edits that helped to improve the clarity of the exposition of the results. M.C. acknowledges support by the European Union (ERC, MMMonsters; grant No. 101117624). This research used resources of the Center for Institutional Research Computing at Washington State University.

*Softwares*: `Numpy` (C. R. Harris et al. 2020); `Matplotlib` (J. D. Hunter 2007); `SciPy` (P. Virtanen et al. 2020); `Astropy` (Astropy Collaboration et al. 2018); `Healpy` (A. Zonca et al. 2019); `Enterprise` (J. A. Ellis et al. 2020).

## ORCID iDs

Niccolò Veronesi https://orcid.org/0000-0001-7678-8218
Maria Charisi https://orcid.org/0000-0003-3579-2522
Stephen R. Taylor https://orcid.org/0000-0003-0264-1453
Jessie Runnoe https://orcid.org/0000-0001-8557-2822
Daniel J. D'Orazio https://orcid.org/0000-0002-1271-6247